\documentstyle[12pt]{article}

\input{psfig}

\begin{document}

\title{Measurements of Charged Current Reactions of $\nu_\mu$ on $^{12}C$}

\author{
L.B. Auerbach,$^8$ R.L. Burman,$^5$ D.O. Caldwell,$^3$ E.D. Church,$^1$ \\
J.B. Donahue,$^5$ A. Fazely,$^7$ G.T. Garvey,$^5$ R.M. Gunasingha,$^7$ 
R. Imlay,$^6$ \\
W.C. Louis,$^5$ R. Majkic,$^{8}$ A. Malik,$^6$ W. Metcalf,$^6$ 
G.B. Mills,$^5$ \\
V. Sandberg,$^5$ D. Smith,$^4$ 
I. Stancu,$^1,$\footnote{Present address: University of Alabama, Tuscaloosa, AL 35487}
M. Sung,$^6$ 
R. Tayloe,$^5,$\footnote{Present address: Indiana University, Bloomington, IN 47405} \\ 
G.J. VanDalen,$^1$ 
W. Vernon,$^2$ N. Wadia,$^6$ D.H. White,$^5$ S. Yellin$^3$\\
(LSND Collaboration) \\
$^1$ University of California, Riverside, Califonia 92521 \\
$^2$ University of California, San Diego, Califonia 92093 \\
$^3$ University of California, Santa Barbara, Califonia 93106 \\
$^4$ Embry Riddle Aeronautical University, Prescott, Arizona 86301 \\
$^5$ Los Alamos National Laboratory, Los Alamos, New Mexico 87545 \\
$^6$ Louisiana State University, Baton Rouge, Louisiana 70803 \\
$^7$ Southern University, Baton Rouge, Louisiana 70813 \\
$^8$ Temple University, Philadelphia, Pennsylvania 19122 }

\date{\today}
\maketitle

\begin{abstract}
Charged current scattering of $\nu_\mu$ on $^{12}C$ has been studied 
using a $\pi^+$ decay-in-flight $\nu_\mu$ beam at the Los Alamos Neutron 
Science Center. 
A sample of $66.9\pm9.1$ events satisfying criteria for 
the exclusive reaction $^{12}C(\nu_\mu,\mu^-)^{12}N_{g.s.}$ was obtained 
using a large liquid scintillator neutrino detector.  
The observed flux-averaged cross section 
$(5.6\pm0.8\pm1.0)\times10^{-41}$ cm$^2$ agrees 
well with reliable theoretical expectations.  
A measurement was also obtained for the inclusive cross section 
to all accessible $^{12}N$ states $^{12}C(\nu_\mu,\mu^-)X$.  
This flux-averaged cross section 
is $(10.6\pm0.3\pm1.8)\times10^{-40}$ cm$^2$ 
which is lower than present theoretical calculations.
\end{abstract}

\clearpage

\section{Introduction}
\label{sec:intro}

Neutrino-nucleus cross sections are needed for the interpretation of 
measurements by many low energy neutrino experiments as well as for 
modeling various astrophysical processes such as supernova explosions.
Low-energy neutrino-nucleus cross sections are also of interest 
because of their role in nuclear structure studies.
The cross sections contain contributions from both axial vector and 
polar vector nuclear currents and thus provide complementary information 
to that provided by eletromagnetic scattering from the nucleus,
which is sensitive only to the nuclear polar vector currents.

Many calculational techniques have been used to determine 
neutrino-nuclear cross sections.
Shell model techniques work best at lower energies 
where transitions to continuum states are not large.
At intermediate energies the Continuum Random Phase Approximation (CRPA) 
is frequently used, while at still higher energies 
the Fermi gas model is thought to work well.
Experimental measurements are, however, necessary 
to establish the range of validity of 
the different calculational techniques.

There are more neutrino cross section measurements for $^{12}C$
than for any other nucleus.
Three experiments, E225\cite{E225} at LAMPF, the KARMEN 
experiment\cite{KARMEN} at the ISIS facility of the Rutherford 
Laboratory and a liquid scintillator neutrino detector 
(LSND)\cite{LSND97a,LSND_nuec01}, have measured both the exclusive 
reaction $^{12}C(\nu_e,e^-)^{12}N_{g.s.}$ and inclusive 
reaction $^{12}C(\nu_e,e^-)^{12}N^*$ to the other accessible 
excited states of $^{12}N$.
In these measurements the $\nu_e$ flux arises from $\mu^+$ decay at rest 
with $E_\nu<52.8$ MeV.
As a result of the low neutrino energy, transitions occur almost entirely 
to a few low lying states of $^{12}N$, and over 60$\%$ of 
the total cross section is to the $^{12}N$ ground state.
The cross section for producing the $^{12}N$ ground state can be 
predicted with an accuracy of $\approx5\%$ by using model-independent 
form factors that can be reliably extracted from other 
measurements\cite{Fukugita88}.
All three experimental measurements of 
the $^{12}C(\nu_e,e^-)^{12}N_{g.s.}$ cross section agree well with the 
expected value.
Calculation of the inclusive cross section to the excited states 
of $^{12}N$ is model dependent and is a less certain procedure.
The Fermi gas model (FGM) is not reliable in this instance 
because the low neutrino energy leads to momentum 
transfers ( $Q<100$ MeV/$c$ ), much smaller 
than the Fermi momentum ( 200 MeV/$c$ ) in carbon.
Thus extensive modeling of the nuclear dynamics is necessary
\cite{Donnelly,Kolbe94,Kolbe99,Volpe00,Hayes00,Auerbach97,Singh98}.
Final LSND results for this reaction\cite{LSND_nuec01} agree 
with a recent shell model calculation \cite{Hayes00}.
They are somewhat lower than, but consistent with, 
a recent CRPA calculation \cite{Kolbe99}.

Measurements also exist for two processes closely related 
to $\nu_e$ carbon scattering: $\mu^-$ capture on $^{12}C$\cite{Suzuki87} 
and $\nu_\mu$ scattering on carbon using a beam of $\nu_\mu$ 
from $\pi^+$ decay-in-flight\cite{LSND95,LSND97b}.
For the $\nu_e$ carbon measurement $\overline{E}_\nu\approx40$ MeV, 
$\overline{Q}\approx50$ MeV/$c$, and 
the inclusive cross section is dominated 
by transitions to low multipoles ($1^+,1^-,2^-$).
In contrast, for the $\nu_\mu$ carbon 
measurement $\overline{E}_\nu\approx170$ MeV, 
$\overline{Q}\approx200$ MeV/$c$, and excitations occur up to 100 MeV.
The $\mu^-$ capture process, which occurs from the $S$ state, 
is intermediate between these two processes 
with $\overline{Q}\approx90$ MeV/$c$.
Because these three processes occur at different energies and 
momentum-transfers 
they constrain different aspects of theoretical calculations.
A challenging test of a theoretical procedure is its ability 
to predict all three processes.
We note that the inclusive cross section is strongly energy 
and momentum transfer dependent.
Thus the flux averaged cross section for the 
reaction $^{12}C(\nu_\mu,\mu^-)^{12}N^*$ is approximately 200 times larger
than the lower energy cross section for $^{12}C(\nu_e,e^-)^{12}N^*$.

The measurement\cite{LSND95,LSND97b} of the inclusive cross section 
for $^{12}C(\nu_\mu,\mu^-)X$ several years ago 
by LSND attracted substantial interest 
because a CRPA calculation\cite{Kolbe94} predicted 
a cross section almost twice as large as that observed.
An improved calculation by the same group\cite{Kolbe99} together 
with an improved calculation of the neutrino energy spectrum and flux 
discussed in Section \ref{sec:source} reduced 
but did not eliminate the discrepancy with the measured cross section.
Recent calculations using the shell model\cite{Volpe00,Hayes00} are 
in better agreement with the measured cross section.
In this paper we present final LSND results for the inclusive cross 
section for $^{12}C(\nu_\mu,\mu^-)X$ and 
for the exclusive reaction $^{12}C(\nu_\mu,\mu^-)^{12}N_{g.s.}$.

\section{The Neutrino Source}
\label{sec:source}

The data reported here were obtained between 1994 and 1998 
at the Los Alamos Neutron Science Center (LANSCE) primarily
using neutrinos produced at the A6 proton beam stop.
As discussed below some neutrinos were also produced at upstream 
targets A1 and A2.  The neutrino source is described in detail
elsewhere\cite{LSND_NIM}.

\begin{figure}
\centerline{\psfig{figure=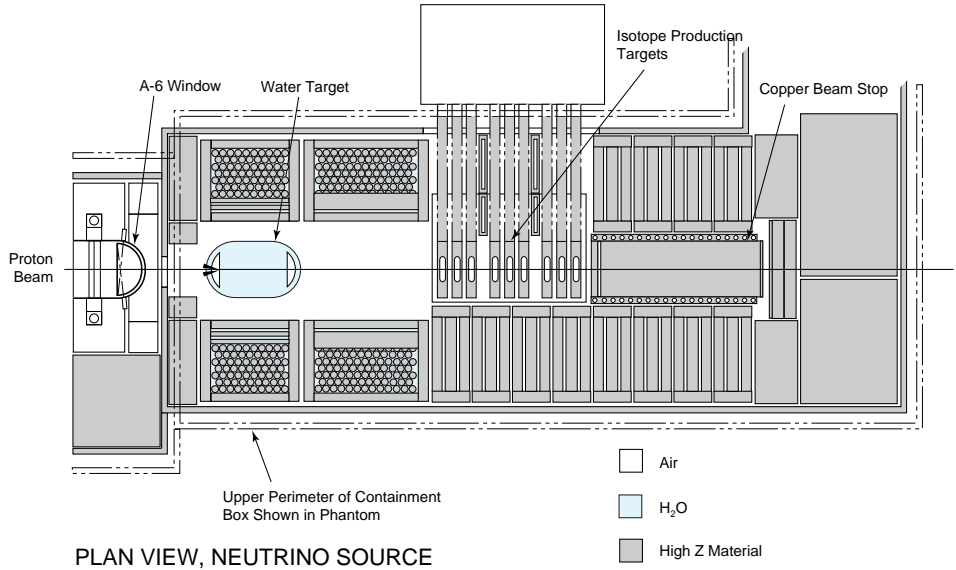,height=3in}}
\caption{The layout of the A6 beam stop, 
as it was configured for the 1993-1995 data taking.}
\label{fig:target}
\end{figure}

In 1994 and 1995 the beam stop consisted of a 30-cm water target  
followed by a 50-cm decay region, isotope production stringers 
and a copper beam dump as shown in Fig. \ref{fig:target}.
The high-intensity 798 MeV proton beam from the linear accelerator 
generates a large pion flux from the water target.
The flux of $\nu_\mu$ and $\bar{\nu}_\mu$ used for the 
measurements reported here arise from the decay in flight (DIF) 
of $\pi^+$ and $\pi^-$. 
For the LANSCE proton beam and beam stop configuration $\pi^+$ production 
exceeds $\pi^-$ production by a factor of approximately 8 and even more 
for high energy pions.  
Approximately 3.4\% of the $\pi^+$ and 5\% of 
the $\pi^-$ decay in flight.  
Upstream targets contributed 6\% to the DIF neutrino flux.  
For the 1995 run, the water target was removed for 32\% of the 7081 C 
of beam.  
For this portion of the run the DIF $\nu_\mu$ flux was reduced 
approximately 50\%.

After the 1995 run the beam stop was substantially modified 
for accelerator production of tritium (APT) tests.
The replacement of the water target by a close-packed, high-Z target 
resulted in reduced $\pi^+$ production largely due to the change in the 
neutron to proton ratio in the target.  
The closer packing of materials for 
the APT stop reduced the fraction of $\pi^\pm$ which decay in flight.  
The resulting $\nu_\mu$ DIF flux per incident proton is only one-half 
of that obtained with the water target.  
There were no upstream targets for almost all of the data 
taken with the APT target in 1996-1998. 

The LANSCE beam dump has been used as the neutrino source for previous 
experiments\cite{Willis80,Krak92,Free93}.  
These experiments primarily used neutrinos from the decay at rest (DAR) 
of stopped $\pi^+$ and $\mu^+$.
A calibration experiment\cite{All89} measured the rate of stopped $\mu^+$ 
from a low-intensity proton beam incident on an instrumented beam stop.
The rate of stopped $\mu^+$ per incident proton was measured 
as a function of several variables and 
used to fine-tune a beam dump simulation program\cite{Bur90}.
The simulation program could then be used to calculate the flux 
for any particular beam dump configuration.
The calibration experiment determined the DAR flux to $\pm7\%$ 
for the proton energies and beam stop configurations used at LANSCE.  
There are greater uncertainties in the DIF fluxes. 
Uncertainties in the energy spectra of the $\pi^\pm$ which decay in
flight lead to uncertainties in both the magnitudes and shapes of the
$\nu_\mu$ and $\bar{\nu}_\mu$ energy spectra.  
The shapes of the $\nu_\mu$ and $\bar{\nu}_\mu$ energy spectra are 
particularly important for the measurement of inclusive cross sections, 
since these cross sections have a strong energy dependence.
We have studied in detail the procedure of Ref. \cite{Bur90} 
for calculating the DIF flux and slightly revised it 
to incorporate more recent $\pi^+$ production 
results\cite{Langenbrunner93} and 
to remove some slight distortions arising from use of finite bins 
in the pion production angle\cite{Imlay98}.
The primary effect is to reduce the $\nu_\mu$ flux above 200 MeV.

The largest uncertainties in the DIF flux arise 
from systematic effects in the calibration experiment (5\%), 
uncertainties in the $\pi^+$ production cross sections used 
in the simulation (10\%) and 
other systematic effects in the simulation (7\%).
For the 1994-1995 data the upstream targets introduced a small additional 
uncertainty in the flux.
For 1996-1998 data the geometry of the beam stop configuration was 
more complicated than that used in 1994-1995.
The uncertainty in the DIF flux for neutrinos 
above muon production threshold is
estimated to produce an uncertainty in the measured cross section of 15\%.
This uncertainty provides the largest source of systematic error for 
the cross sections presented here.

The LANSCE proton beam typically had a current of 800 $\mu$A 
at the beam stop during the 1994-1995 running period 
and 1000 $\mu$A for 1996-1998. 
For 1994 and 1995 the energy was approximately 770 MeV at the beam stop 
due to energy loss in upstream targets, 
while it was approximately 800 MeV in 1996, 1997 and 1998.

\begin{figure}
\centerline{\psfig{figure=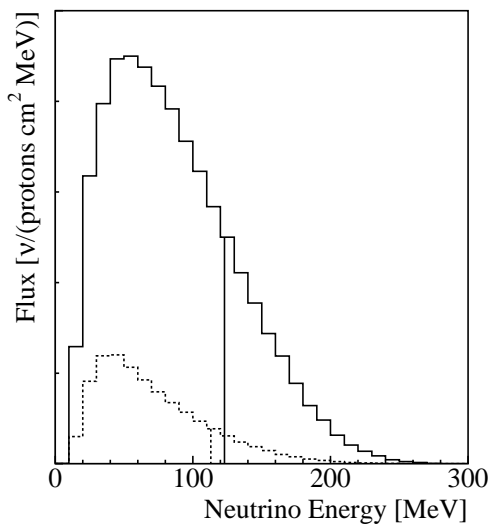,height=3in}}
\caption{The solid line shows the flux shape of $\nu_\mu$ from 
$\pi^+$ decay-in-flight.
The dashed line shows the $\bar{\nu}_\mu$ flux 
from $\pi^-$ decay-in-flight for the same integrated proton beam.
The muon production threshold energy for each spectrum is shown by 
a vertical line.}
\label{fig:dif_flux}
\end{figure}

Table \ref{ta:flux} shows for each year the calculated $\nu_\mu$ flux 
above the threshold (123.1 MeV) for muon production on carbon and 
averaged over the LSND detector.
\begin{table}
\centering
\caption{The $\nu_\mu$ and $\bar{\nu}_\mu$ fluxes  
above the threshold energy for muon production and 
averaged over the LSND detector.}
\begin{tabular}{ccc}
\hline
Year~&~~$\nu_\mu$ flux[cm$^{-2}$]~~&~~$\bar{\nu}_\mu$ flux[cm$^{-2}$]~~\\
\hline
1994 &     $6.04\times10^{11}$     &        $6.06\times10^{10}$        \\
1995 &     $5.97\times10^{11}$     &        $6.21\times10^{10}$        \\
1996 &     $2.06\times10^{11}$     &        $2.91\times10^{10}$        \\
1997 &     $4.46\times10^{11}$     &        $5.86\times10^{10}$        \\
1998 &     $1.74\times10^{11}$     &        $2.32\times10^{10}$        \\
\hline
Total &    $2.03\times10^{12}$     &        $2.34\times10^{11}$        \\
\hline
\end{tabular}
\label{ta:flux}
\end{table}
The $\bar{\nu}_\mu$ flux above threshold (113.1 MeV) for the process 
$\bar{\nu}_\mu + p \rightarrow \mu^+ + n$ is also shown.  
Figure \ref{fig:dif_flux} shows the calculated $\nu_\mu$ 
and $\bar{\nu}_\mu$ energy spectra.

\section{The LSND Detector}
\label{sec:lsnd}

\begin{figure}
\centerline{\psfig{figure=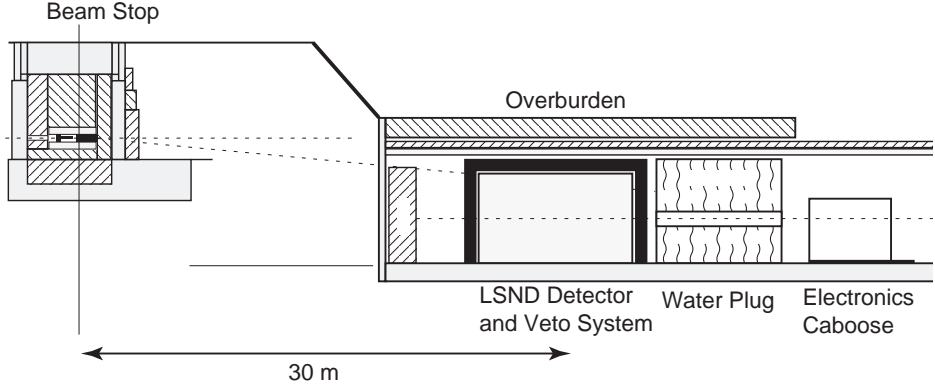,height=2in}}
\caption{Detector enclosure and target area configuration, 
elevation view.}
\label{fig:detector}
\end{figure}

The detector is located 29.8 m downstream of the proton beam stop 
at an angle of $12^\circ$ to the proton beam.  
Figure \ref{fig:detector} shows a side-view of the setup.  
Approximately 2000 g/cm$^2$ of shielding above the detector attenuates 
the hadronic component of cosmic rays to a negligible level.  
The detector is also well shielded from the beam stop 
so that beam-associated neutrons are attenuated to a negligible level.  
Enclosing the detector, except on the bottom, 
is a highly efficient liquid scintillator veto shield 
which is essential to reduce contributions 
from the cosmic ray muon background to a low level.    
Reference \cite{LSND_NIM} provides a detailed description of the detector,
veto, and data acquisition system (DAQ) which we briefly review here.  

The detector is an approximately cylindrical tank containing 167 
tons of liquid scintillator and viewed by 1220 uniformly spaced $8''$ 
Hamamatsu photomultiplier tubes (PMT) covering $25\%$ of the surface 
inside the tank wall. 
When the deposited energy in the tank exceeds a threshold of 
approximately 4 MeV electron-equivalent energy and 
there are fewer than 4 PMT hits in the veto shield, 
the digitized time and pulse height of each of these PMTs 
(and of each of the 292 veto shield PMTs) are recorded.  
A veto, imposed for 15.2 $\mu$s following the firing of $>5$ veto PMTs, 
substantially reduces ($10^{-3}$) 
the large number of background events arising from the decay 
of cosmic ray muons that stop in the detector.  
Activity in the detector or veto shield during the 51.2 $\mu$s 
preceding a primary trigger is also recorded, 
provided there are $>17$ detector PMT hits or $>5$ veto PMT hits.  
This activity information is used in the analysis to identify events 
arising from muon decay. 
In particular, in this analysis the activity information is used 
to identify $\mu^-$ from the reaction $\nu_\mu+^{12}C\rightarrow\mu^-+X$.
For such events the $e^-$ from the subsequent 
decay $\mu^-\rightarrow e^-+\nu_\mu+\bar{\nu}_e$ provides 
the primary trigger.  
Activities with $>3$ veto PMT hits mostly arise from cosmic ray muons 
and are rejected in the analysis.  
It should also be noted that the 15.2 $\mu$s veto applies only to 
the primary trigger and not to the activities preceding a valid trigger.
Data after the primary event are recorded for 1 ms with a threshold 
of 21 PMTs (approximately 0.7 MeV electron-equivalent energy).  
This low threshold is used for the detection of 2.2 MeV $\gamma$ 
from neutron capture on free protons.  
Muon events with associated neutrons arise from the processes 
$\bar{\nu}_\mu p\rightarrow\mu^+n$, $\bar{\nu}_\mu C\rightarrow\mu^+nX$, 
and $\nu_\mu C\rightarrow\mu^-nX$.

The detector operates without reference to the beam spill, 
but the state of the beam is recorded with the event.  
Approximately 94\% of the data is taken between beam spills.  
This allows an accurate measurement and subtraction of cosmic ray 
background surviving the event selection criteria. 

The detector scintillator consists of mineral oil ($CH_2$) in which is 
dissolved a small concentration (0.031 g/l) of b-PBD\cite{Ree93}. 
This mixture allows the separation of \v{C}erenkov light and 
scintillation light and produces about 33 photoelectrons per MeV of 
electron energy deposited in the oil.  
The combination of the two sources of light provides direction 
information for relativistic particles and makes 
particle identification (PID) possible.
Note that the oil consists almost entirely of carbon and hydrogen.  
Isotopically the carbon is $1.1\%~^{13}C$ and $98.9\%~^{12}C$. 
Stopping $\mu^-$ are captured on $^{12}C$ 8\% of the time in the LSND 
detector.  The $\mu^\pm$ which decay are readily identified as muons 
by the presence of subsequent spatially correlated Michel electrons.

The veto shield encloses the detector on all sides except the bottom.  
Additional counters were placed below the veto shield 
after the 1993 run to reduce cosmic ray background entering 
through the bottom support structure.   
More counters were added after the 1995 run.  
The main veto shield\cite{Nap89} consists of a 15-cm layer of 
liquid scintillator in an external tank and 15 cm of lead shot 
in an internal tank.  
This combination of active and passive shielding tags cosmic ray muons 
that stop in the lead shot.  
A veto inefficiency $<10^{-5}$ is achieved with 
this detector for incident charged particles.

\section{Analysis Techniques}
\label{sec:analysis}

In the analysis presented in this paper we require a $\mu^\pm$ followed 
by a delayed coincidence with a decay $e^\pm$.  
As a result of this coincidence requirement a clean beam excess sample 
of events can be obtained with relatively loose selection criteria.  
Furthermore, it is easy to verify that the events in this sample arise 
from muon decay since the muon lifetime and 
the decay electron energy spectrum are well known.

Each event is reconstructed using the hit time and pulse height 
of all hit PMTs in the detector\cite{LSND_NIM}.
The present analysis relies on the reconstructed energy, position, 
and two PID parameters, $\chi'_{tot}$ and $\alpha$, 
as described later in this section.
The parameters $\chi'_{tot}$ and $\alpha$ are used to distinguish 
electron events from events arising from interactions of 
cosmic ray neutrons in the detector.
Fortunately, it is possible to directly measure the response 
of the detector to electrons and neutrons in the energy range of interest 
for this analysis by using copious control data samples.
We also use a Monte Carlo simulation, LSNDMC\cite{McI95}, 
to simulate events in the detector using GEANT.

The response of the detector to electrons was determined from a large, 
essentially pure sample of electrons (and positrons) from the decay of 
stopped cosmic ray $\mu^\pm$ in the detector.
The known energy spectra for electrons from muon decay was used 
to determine the absolute energy calibration, 
including its small variation over the volume of the detector.
The energy resolution was determined from the shape of the electron 
energy spectrum and was found to be 6.6\% at the 52.8 MeV end-point.  

For relativistic electrons in the LSND detector approximately 65\% 
of the photoelectrons arise from direct or reradiated \v{C}erenkov light 
and only 35\% from scintillator light. 
For muons, the threshold kinetic energy for \v{C}erenkov radiation 
in the LSND detector is 39 MeV.  
For the sample of muons analyzed in this paper only about half 
are above \v{C}erenkov threshold and none fully relativistic.  
As a result, the light output per MeV of energy loss for the 
muons is significantly less than that for relativistic electrons.  
There is no calibration sample available of low-energy muons 
with known energies.  
Thus we rely on the Monte Carlo simulation LSNDMC for muons.  
We discuss the muon energy scale further in Sections \ref{sec:gstate} 
and \ref{sec:inclusive} 
when we compare observed and expected energy distributions. 

There are no tracking devices in the LSND detector. 
Thus, event positions must be determined solely from the PMT information.
The reconstruction process determines an event position by minimizing 
a function $\chi_r$ which is based on the time of each PMT hit corrected 
for the travel time of light from the assumed event position 
to the PMT\cite{LSND_NIM}.
The procedure used in several previous analyses systematically shifted 
event positions away from the center of the detector and thus 
effectively reduced the fiducial volume\cite{LSND97a,At96}.
The reconstruction procedure has been analyzed in detail and 
an improved reconstruction procedure was developed which reduces this 
systematic shift and provides substantially better position resolution.
This procedure also provides results which agree well with positions 
obtained from the event likelihood procedure described 
in Ref \cite{At98}. 
In the analysis presented in this paper, a fiducial cut is imposed 
by requiring $D>35$ cm, where $D$ is the distance between 
the reconstructed electron position and the surface tangent 
to the faces of the PMTs. 
For the muon we require $D>0$ cm.

The particle identification procedure is designed to separate particles 
with velocities well above \v{C}erenkov threshold from particles 
below \v{C}erenkov threshold.
The procedure makes use of the four parameters defined 
in Refs. \cite{LSND_NIM}.
Briefly, $\chi_r$ and $\chi_a$ are the quantities minimized for 
the determination of the event position and direction, $\chi_t$ is 
the fraction of PMT hits that occur more than 12 ns after the fitted 
event time and $\chi_{tot}$ is proportional to the product of $\chi_r$, 
$\chi_a$ and $\chi_t$.

Several previous LSND analyses\cite{LSND97a,LSND97b,At96} have 
used $\chi_{tot}$ for particle identification.
The distribution of $\chi_{tot}$ for electrons, however, has 
a small variation with electron energy and with the position of the event.
Therefore, in this paper, we used a modified variable, $\chi'_{tot}$, 
with approximately a mean of zero and sigma of one, 
independent of the electron energy and positions. 
We also used the variable, $\alpha$, which is based on the event 
likelihood procedures discussed in Ref. \cite{At98}. 
It is similar to the parameter $\rho$ discussed there, 
which is based on the ratio of \v{C}erenkov to scintillator light.
The $\alpha$ parameter varies from 0 to 1 and 
peaks at one for electrons and at 0.3 for neutrons.
As discussed in Refs. \cite{LSND_nuec01} and \cite{lsnd_nuee01}, 
the combination $\chi_\alpha=\chi_{tot}^\prime+10(1-\alpha)$ provides 
better separation of electrons, muons, and neutrons 
than $\chi_{tot}^\prime$ by itself.

We note that a modest particle identification requirement was imposed 
in the initial data processing that created the samples analyzed here. 
The effect of this requirement is also included in the analysis.

Cosmic ray background which remains after all selection criteria 
have been applied is well measured with the beam-off data and 
subtracted using the duty ratio, the ratio of beam-on time 
to beam-off time.
The subtraction was performed separately for each year's data 
using the measured duty ratio for that year.
The ratio averaged over the full data sample was 0.0632.
In the present analysis the beam-off background is very small($<2\%$) 
because we require a low energy muon that is well correlated in space 
and time with a Michel electron.

\section{Event Selection Criteria}
\label{sec:electron}

The analysis is designed to select the $\mu^-$ from the reaction 
$\nu_\mu+^{12}C\rightarrow \mu^-+X$ and the subsequent electron from 
the decay $\mu^-\rightarrow e^-+\bar{\nu}_e+\nu_\mu$.  
In the LSND detector medium 92\% of the stopped $\mu^-$ decay 
and 8\% are captured.  
The $\mu^-$ and other particles arising from the charge-changing neutrino 
interaction produce light that causes an average of 250 PMTs to fire.  
The detector charge $Q_\mu$, measured in photoelectrons, arises mostly 
from the $\mu^-$ but includes contributions from other particles 
produced in the reaction such as protons and $\gamma$'s.

\begin{figure}
\centerline{\psfig{figure=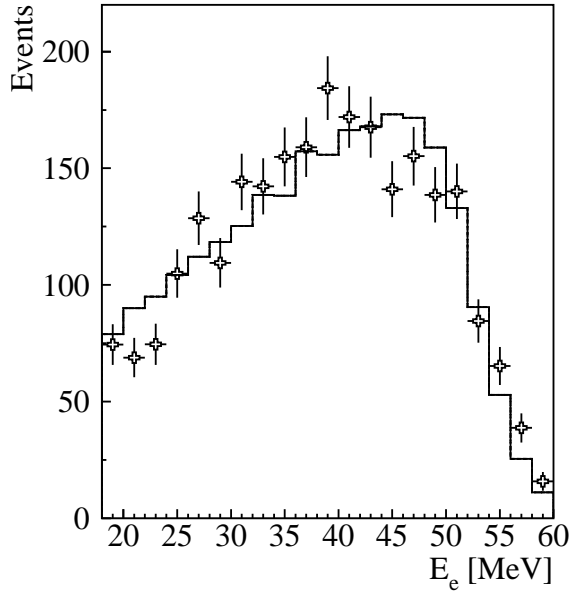,height=3.5in}}
\caption{The observed energy distribution of electrons from 
$\mu^-$ decay for the inclusive sample, $^{12}C(\nu_\mu,\mu^-)X$. 
The histogram shows the expected energy distribution of Michel electrons 
from Monte Carlo simulation.}
\label{fig:Eelec}
\end{figure}

\begin{figure}
\centerline{\psfig{figure=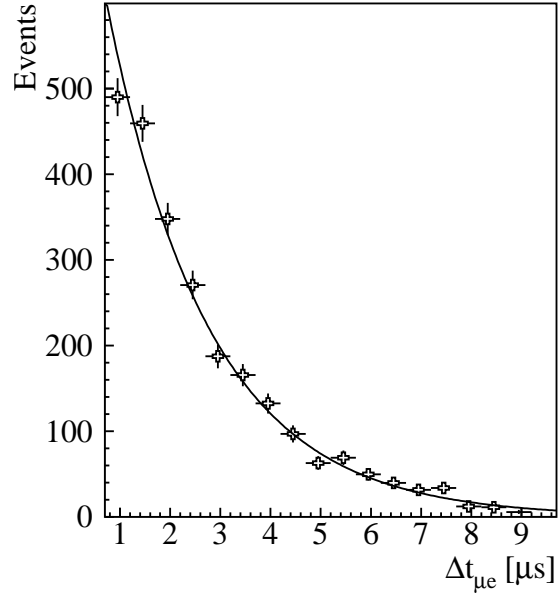,height=3.5in}}
\caption{The distribution of the time difference, $\Delta t_{\mu e}$, 
between the $\mu^-$ and the decay $e^-$ in the inclusive 
sample, $^{12}C(\nu_\mu,\mu^-)X$. 
The best fit (solid line) curve corresponds to a lifetime 
of $2.03\pm0.05~\mu$s.}
\label{fig:Temu}
\end{figure}

\begin{figure}
\centerline{\psfig{figure=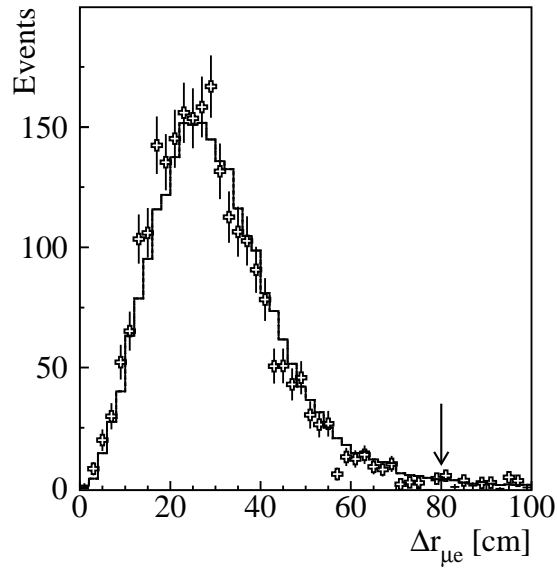,height=3.5in}}
\caption{The distribution of the distance between the reconstructed 
positions of the $\mu^-$ and $e^-$ 
in the beam-excess inclusive sample, $^{12}C(\nu_\mu,\mu^-)X$. 
The histogram is the prediction from the Monte Carlo simulation, 
normalized to the data.}
\label{fig:Demu}
\end{figure}

Tables \ref{ta:electron} and \ref{ta:muon} respectively show the 
selection criteria and corresponding efficiencies 
for the electron and for the muon.  
\begin{table}
\centering
\caption{The electron selection criteria and corresponding efficiencies.}
\begin{tabular}{ccc}
\hline
   Quantity            &    Criteria           &   Efficiency    \\    
\hline
Fiducial volume        & $D>35$ cm,            & 0.880$\pm$0.055 \\
Electron energy        & $18<E_e<60$ MeV       & 0.924$\pm$0.010 \\
Particle ID            & $\chi_\alpha<4$       & 0.940$\pm$0.018 \\
In-time veto           & $<4$ PMTs             & 0.986$\pm$0.010 \\
Future activity        & $\Delta t_f>8.8~\mu$s & 0.991$\pm$0.003 \\
Past activity          & See text              & 0.858$\pm$0.010 \\
Trigger veto        & $> 15.1 \mu$s         & 0.760$\pm$0.010 \\
DAQ dead time       &                     & 0.977$\pm$0.010 \\
Tape dead time      &                    & 0.981$\pm$0.010 \\
\hline
Total               &                       & 0.467$\pm$0.032 \\
\hline
\end{tabular}
\label{ta:electron}
\end{table}
\begin{table}
\centering
\caption{The muon selection criteria and corresponding efficiencies.}
\begin{tabular}{ccc}
\hline
Quantity                      & Criteria              & Efficiency      \\
\hline
Fiducial volume               & $D>0$ cm              & 0.978$\pm$0.010 \\
Not $\mu^-$ Capture           &      --               & 0.922$\pm$0.005 \\
Muon Energy ($e^-$ equivalent)& $E<70$ MeV            & 0.990$\pm$0.008 \\
Spatial Correlation           & $\Delta r<80$ cm      & 0.990$\pm$0.002 \\
$\mu$ Decay Time              & $0.7<\Delta t<9~\mu$s & 0.687$\pm$0.005 \\
Intime Veto                   & $< 4$ PMTs            & 0.988$\pm$0.010 \\
\hline
Total                         &                       & 0.599$\pm$0.011 \\
\hline
\end{tabular}
\label{ta:muon}
\end{table}
For events in the decay-in-flight sample the event position is 
best determined from the reconstructed electron position rather 
than the reconstructed muon position, especially for events 
with low-energy muons.  
Therefore, the fiducial selection is imposed primarily on the electron.  
The reconstructed electron is required to be a distance $D>35$ cm 
from the surface tangent to the faces of the PMTs.  
There are $3.65\times10^{30}~^{12}C$ nuclei within this fiducial volume. 
The fiducial volume efficiency, defined to be the ratio of the number 
of events reconstructed within the fiducial volume to the actual number 
within this volume, was determined to be $0.880\pm0.055$.
This efficiency is less than one because there is a systematic shift 
of reconstructed event positions away from the center of the detector, 
as discussed in Section \ref{sec:analysis}.  
The muon is required to reconstruct only inside the region $D>0$ cm.  
A lower limit on the electron energy of 18 MeV eliminates 
the large background from $^{12}B$ $\beta$ decay 
created by the capture of cosmic ray $\mu^-$ on $^{12}C$.  
Figure \ref{fig:Eelec} shows the observed electron energy distribution 
compared with the expected energy distribution of Michel electrons 
from Monte Carlo simulation.
The distribution of the time, $\Delta t_{\mu e}$, between the muon and 
electron candidates, shown in Fig. \ref{fig:Temu}, agrees well 
with the 2.03 $\mu$s $\mu^-$ lifetime in mineral oil. 
The best fit, also shown, corresponds to a lifetime 
of $2.03\pm0.05~\mu$s. 
The requirement $\Delta t_{\mu,e}\geq0.7\mu$s is imposed 
to insure that the $\mu$ and $e$ are clearly separated in 
the trigger and in the readout of the data.  
The excellent agreement with expectations in Figs. \ref{fig:Eelec} 
and \ref{fig:Temu} clearly shows that the events arise from muon decay.  
There is an 8\% loss of events due to $\mu^-$ capture 
in the detector medium.  
Figure \ref{fig:Demu} shows the spatial separation $\Delta r$ between 
the reconstructed muon and electron positions.  
A loose requirement, $\Delta r<0.8$ m, is imposed to minimize 
the background from accidental $\mu,e$ 
correlations while retaining high acceptance.  

Many of the selection criteria are designed to reduce the cosmic ray 
background, especially that arising from the decay of cosmic ray muons 
which stop in the detector. 
Both the muon and the electron candidates are required 
to have fewer than 4 PMT hits in the veto and no bottom counter 
coincidence during the 500 ns event window.  
The detector PMT faces are 25 cm inside the tank and 
thus stopping cosmic ray muons must traverse at least 60 cm of 
oil to reach the fiducial volume.  
As a result, these muons typically produce a large detector signal.  
The requirement $E<$70 MeV, where $E$ is the electron equivalent energy 
of the muon, eliminates most such background events with almost no loss 
of acceptance for muons arising from neutrino interactions. 

Muons which are misidentified as electrons are removed by requiring 
that there be no future activity consistent with a Michel electron.
Any electron candidate with future activity 
with fewer than 4 veto PMT hits and 
more than 50 detector PMT hits within 8.8 $\mu$s is rejected.

Frequently, in addition to the candidate muon 
which satisfies the criteria in Table \ref{ta:muon}, 
there are one or more other activities prior to the electron.  
If an activity is due to a stopping muon, 
that muon could be the parent of the observed electron.  
Therefore an event is rejected if, in the 35 $\mu$s interval 
prior to the electron, there is an activity (other than the muon) 
with $Q>3000$ pe or an activity with $>4$ PMT hits in the veto 
and $>100$ PMT hits in the detector. 
We also reject any event with a past activity within 51 $\mu$s 
with $>5$ veto PMT hits and $>500$ detector PMT hits.  
A cut is performed during initial data processing on past activities 
that are spatially correlated with the primary event, 
within 30 $\mu$s of the primary event, and have $\geq 4$ veto PMT hits.  

The acceptances for the past activity and in-time veto cuts 
are obtained by applying these cuts to a large sample of events 
triggered with the laser used for detector calibration.  
These laser events are spread uniformly through the run and 
thus average over the small variation in run conditions.  

\begin{figure}
\centerline{\psfig{figure=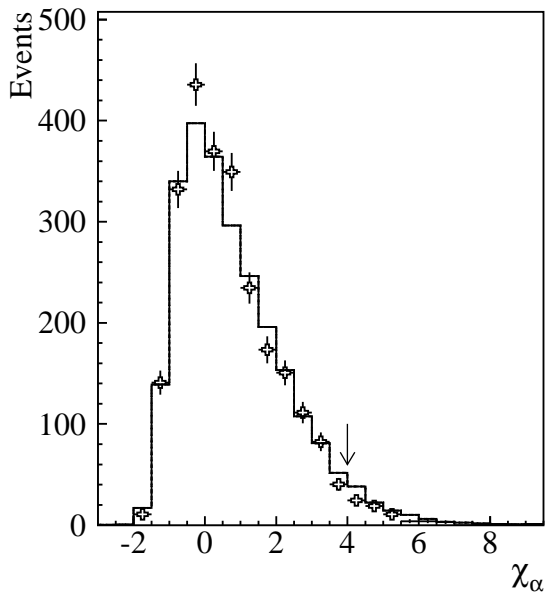,height=3.5in}}
\caption{The distribution of the particle identification 
parameter $\chi_\alpha$ of electrons from $\mu^-$ decay 
for the inclusive sample $^{12}C(\nu_\mu,\mu^-)X$. 
The histogram shows the $\chi_\alpha$ distribution of Michel electrons.}
\label{fig:pid}
\end{figure}

Only a loose particle ID requirement, $\chi_\alpha<4.0$, was imposed 
on the electron and none on the muon.  
A sample of Michel electrons (electrons from the decay 
of stopped $\mu^\pm$) was analyzed to obtain the acceptance of electrons 
for the $\chi_\alpha$ particle identification cut, 
as shown in Fig. \ref{fig:pid}.

\begin{figure}
\centerline{\psfig{figure=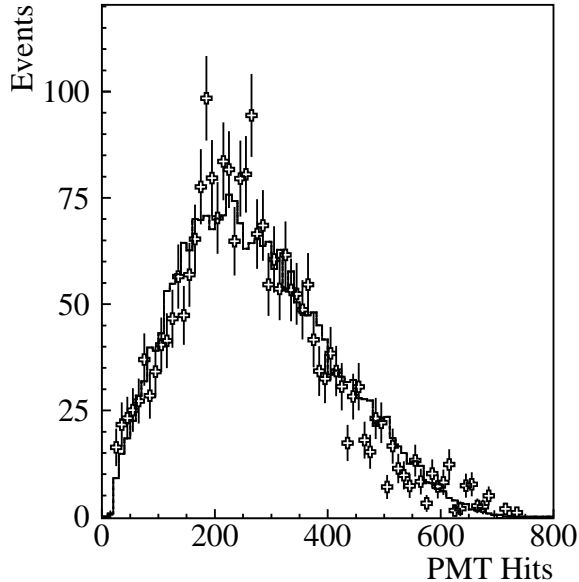,height=3.5in}}
\caption{The observed PMT hit distribution for 
the decay-in-flight sample (including $\nu_\mu+C\rightarrow \mu^-+X$, 
$\bar{\nu}_\mu C\rightarrow\mu^+X$ and 
$\bar{\nu}_\mu p\rightarrow\mu^+n$).
The histogram is the prediction from the Monte Carlo simulation, 
normalized to the data.}
\label{fig:thit}
\end{figure}

The Monte Carlo simulation LSNDMC was used to obtain the PMT hit 
distributions expected from the various processes that contribute to the 
inclusive sample and to the exclusive sample with an identified $\beta$ 
decay.  
Figure \ref{fig:thit} compares the observed and expected distribution 
of PMT hits for muons from inclusive events.  
There is excellent agreement, and thus we expect that the simulation 
provides a reliable estimate of the fraction of events lost 
because they are below the activity threshold of 18 PMT hits 
(roughly 4 MeV).  
For the inclusive sample (exclusive sample) 
we find that only 1.2\%(2.4\%) of the events have fewer than 18 PMT hits.

\section{The Transition to the $^{12}N$ Ground State}
\label{sec:gstate}

\begin{figure}
\centerline{\psfig{figure=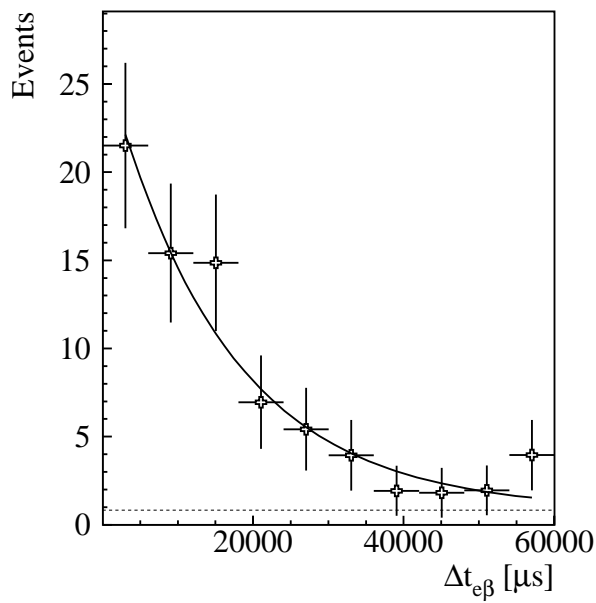,height=3.5in}}
\caption{The distribution of time differences between the electrons and 
$\beta$ in the exclusive sample of $^{12}C(\nu_\mu,\mu^-)^{12}N_{g.s.}$ 
is compared with the expected $\beta$ lifetime. 
The dotted line shows the calculated accidental contribution.}
\label{fig:fdt}
\end{figure}

\begin{figure}
\centerline{\psfig{figure=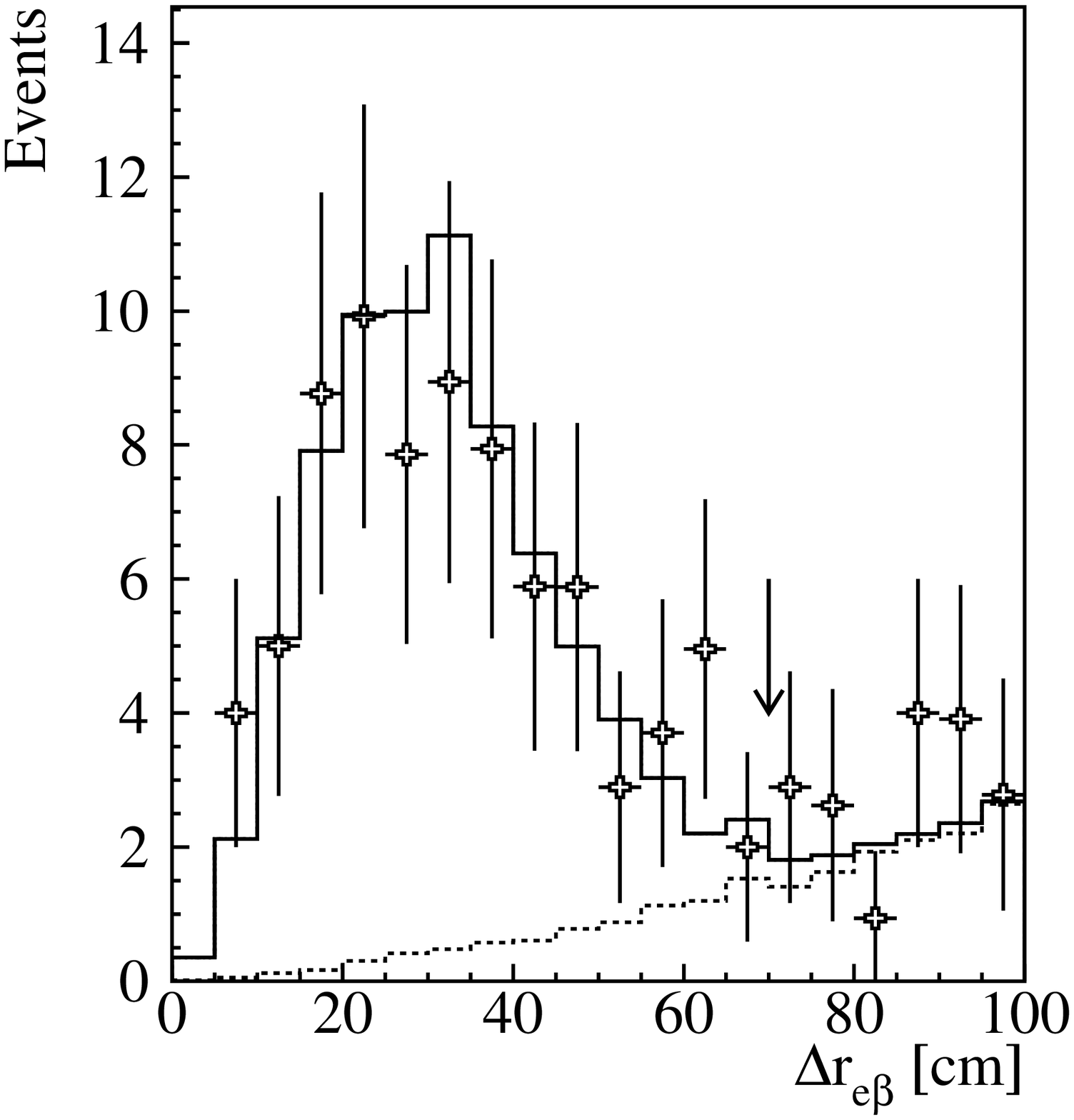,height=3.5in}}
\caption{The distribution of the distances between the electrons and 
$\beta$ for beam-excess events in the exclusive sample 
of $^{12}C(\nu_\mu,\mu^-)^{12}N_{g.s.}$. 
The dashed line shows the estimated accidental contribution.
The solid line shows the expected shape (including the accidental 
contribution) from the Monte Carlo simulation, normalized to data.}
\label{fig:fdr}
\end{figure}

\begin{figure}
\centerline{\psfig{figure=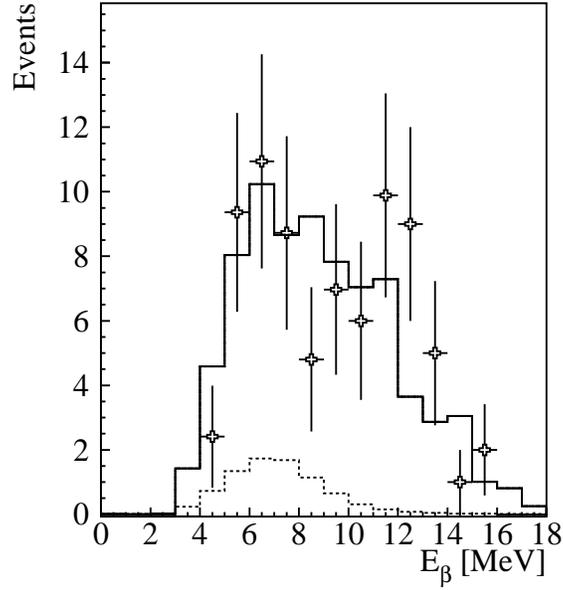,height=3.5in}}
\caption{The distribution of $\beta$ energy 
from the exclusive sample of $^{12}C(\nu_\mu,\mu^-)^{12}N_{g.s.}$. 
The dashed line shows the estimated accidental contribution.
The solid line shows the expected shape (including the accidental 
contribution) from the Monte Carlo simulation, normalized to data.}
\label{fig:ebeta}
\end{figure}

\begin{figure}
\centerline{\psfig{figure=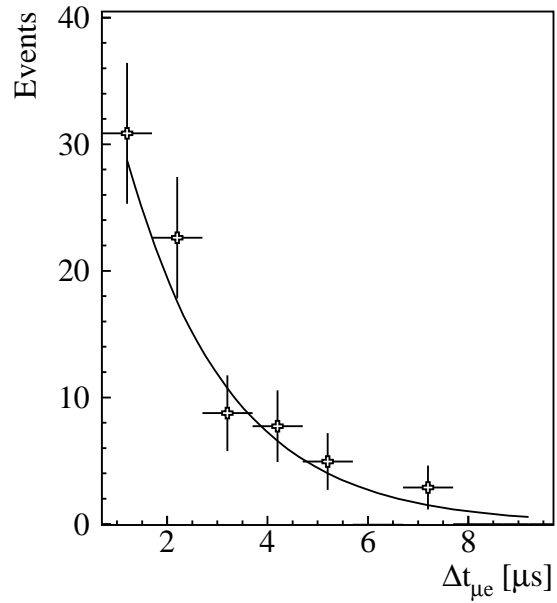,height=3.5in}}
\caption{The distribution of muon decay time
from the exclusive sample of $^{12}C(\nu_\mu,\mu^-)^{12}N_{g.s.}$. 
The curve shows the distribution from the expected muon lifetime 
in the oil.} 
\label{fig:pdt_gs}
\end{figure}

For analysis of the exclusive 
process $^{12}C(\nu_\mu,\mu^-)^{12}N_{g.s.}$ we also require detection 
of the $e^+$ from the $\beta$ decay of $^{12}N_{g.s.}$.  
Therefore, for these events three particles are detected:  
the muon, the decay electron and the positron from the $\beta$ decay 
of $^{12}N_{g.s.}$.  
Table \ref{ta:betaeff} gives the selection criteria and efficiencies for 
the $^{12}N$ $\beta$ decay positron.  
\begin{table}
\centering
\caption{Criteria to select $e^+$ from $N_{g.s.}$ beta decay and 
corresponding 
efficiencies for the reaction $^{12}C(\nu_\mu,\mu^-)^{12}N_{g.s.}$.}
\begin{tabular}{ccc}
\hline
Quantity            & Criteria            &    Efficiency   \\
\hline
$\beta$ decay time  & 52 $\mu$s$<t<60$ ms & 0.974$\pm$0.002 \\ 
Spatial correlation & $\Delta r<0.7$ m    & 0.992$\pm$0.008 \\
PMT threshold       & $>100$ for 1994,    & 0.856$\pm$0.011 \\
                    & $>75$ after 1994    &                 \\
Positron energy     & $E_\beta<18$ MeV    & 0.999$\pm$0.001 \\
Fiducial volume     & $D>0$ cm            & 0.986$\pm$0.010 \\
Trigger veto        & $>15.1~\mu$s        & 0.760$\pm$0.010 \\
In-time veto        & $<4$ PMTs           & 0.988$\pm$0.010 \\
DAQ dead time       &                     & 0.977$\pm$0.010 \\
\hline
Total               &                     & 0.598$\pm$0.016 \\ 
\hline
\end{tabular}
\label{ta:betaeff}
\end{table}
These are the same criteria used previously in an analysis 
of a much larger sample from the analogous 
process $^{12}C(\nu_{e},e^-)^{12}N_{g.s.}$\cite{LSND97a,LSND_nuec01}.  
The $\beta$ decay has a mean lifetime of 15.9 ms and 
maximum positron kinetic energy of 16.3 MeV\cite{FAS90}. 
Figure \ref{fig:fdt} shows the observed $\beta$ decay time distribution 
compared with the expected 15.9 ms lifetime.  
Figure \ref{fig:fdr} shows the distance between the reconstructed 
electron and positron positions for the beam-excess sample.  
A cut was applied at 70 cm, resulting in an acceptance 
of $(99.2\pm0.8)\%$. 
The positron is required to be spatially correlated with the electron 
rather than the muon because the position of the electron in general 
is better determined.  
Following a muon produced by a neutrino interaction, an 
uncorrelated particle, such as the positron from $^{12}B~\beta$ decay, 
will occasionally satisfy all the positron criteria 
including the requirements of time (60 ms) and spatial (70 cm) 
correlation with the electron.  
The probability of such an accidental coincidence can be precisely 
measured from the Michel electron sample.  
The background from this source is also shown in Figs. \ref{fig:fdt} 
and \ref{fig:fdr}.  
The efficiency of 76.0\% caused by the 15.2 $\mu$s veto and the 
trigger dead time of 2.3\% are the same as for the electron.  
Positrons with four or more in-time veto hits 
or any bottom veto coincidence are rejected. 
The Monte Carlo simulation was used to generate the expected distribution 
for the positron energy.  
There was a trigger requirement of 100 PMT hits for 1994 
and 75 PMT hits for 1995-1998.  
The positron was required to have an energy less than 18 MeV.  
Figure \ref{fig:ebeta} compares the observed and expected positron 
energy distributions.  
Figure \ref{fig:pdt_gs} compares the observed and expected 
distributions of muon decay time.  

\begin{figure}
\centerline{\psfig{figure=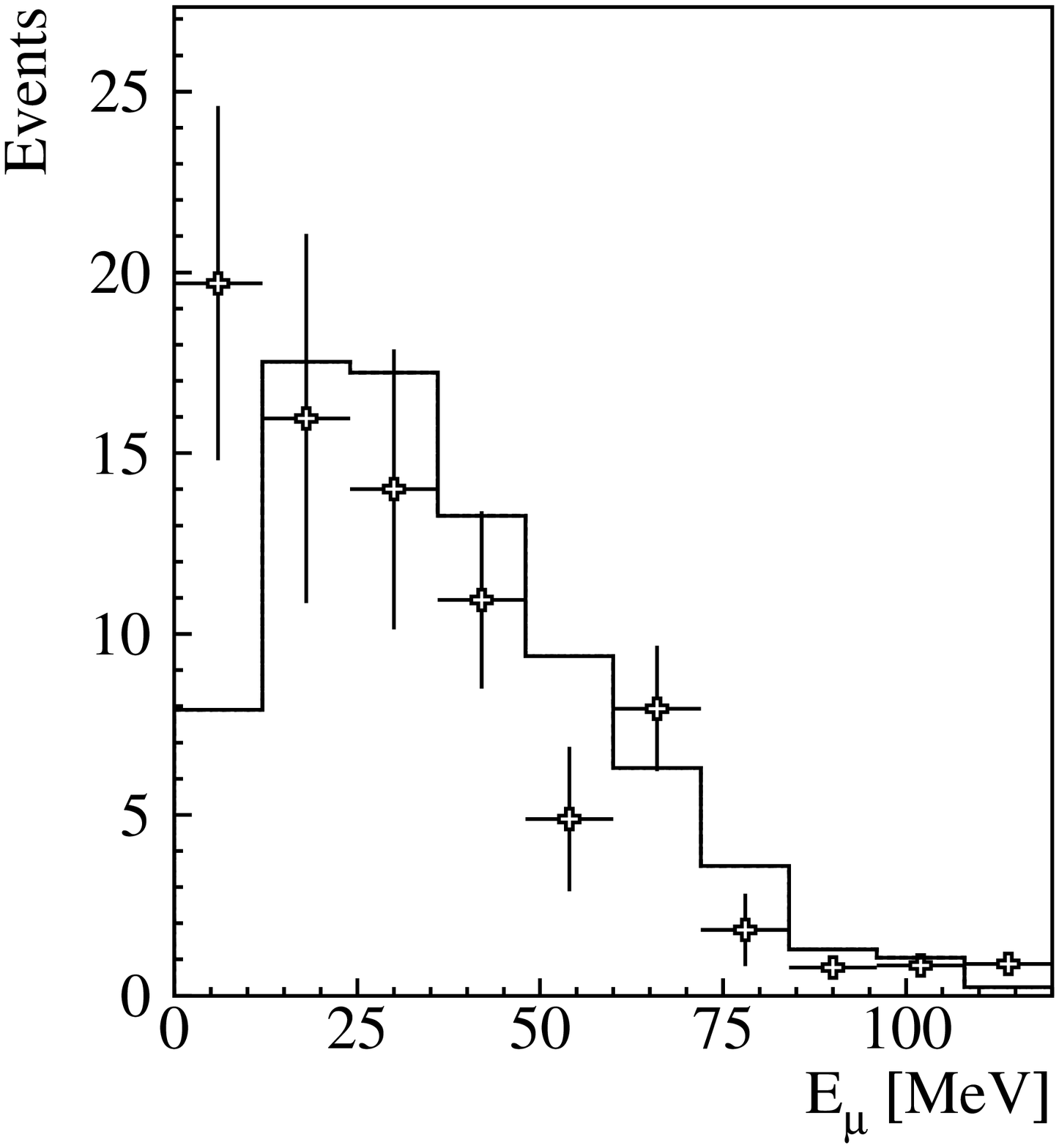,height=3.5in}}
\caption{The observed and expected (solid line) $\mu^-$ kinetic 
energy distribution for beam excess-events 
in the $^{12}C(\nu_\mu,\mu^-)^{12}N_{g.s.}$ sample.}
\label{fig:emu_gs}
\end{figure}

Excited states of $^{12}N$ decay by prompt proton emission and 
thus do not feed down  to the $^{12}N$ ground state or contribute 
to the delayed coincidence rate.  
The form factors required to calculate the cross section are 
well known from a variety of previous measurements\cite{Fukugita88}. 
This cross section and the known $\nu_\mu$ flux are used 
to obtain the expected muon kinetic energy spectrum 
which is compared with the data in Fig.  \ref{fig:emu_gs}. 

As stated in Section \ref{sec:analysis} the energy calibration for muons 
(the conversion from photoelectrons to MeV) is obtained 
from the Monte Carlo simulation LSNDMC.  
For this ground state reaction, the expected muon energy distribution 
should be very reliable. 
Thus the agreement seen in Fig. \ref{fig:emu_gs} provides confirmation 
for the muon energy calibration within the limited statistics. 

There are two sources of background.  
The largest arises from the accidental coincidence of a positron 
candidate with an event from the inclusive sample 
of neutrino-induced muons.  
The probability of an uncorrelated particle 
satisfying all the positron criteria, including the requirements of time 
(60 ms) and spatial correlation (70 cm) with the electron, 
can be precisely measured from a large Michel electron sample.  
The second background arises from the 
process $^{12}C(\bar{\nu}_\mu,\mu^+)^{12}B_{g.s.}$, 
where we detect the $e^-$ from the $\beta$ decay 
of the $^{12}B$ ground state\cite{Engel96}. 
This background is small primarily because the flux of high-energy 
$\bar{\nu}_\mu$ is approximately a factor of 10 lower than 
the corresponding $\nu_{\mu}$ flux and because the $^{12}B_{g.s.}$ 
lifetime is longer than the $^{12}N_{g.s.}$ lifetime.

Table \ref{ta:gstate} shows the number of beam excess events, 
the number of background events, the efficiency, the neutrino flux 
for $E_\nu>123.1$ MeV, and the cross section averaged over the flux.  
\begin{table}
\centering
\caption{Beam-excess events, background, efficiency, neutrino flux and 
flux-averaged cross section for the exclusive 
reaction $^{12}C(\nu_\mu,\mu^-)^{12}N_{g.s.}$.}
\begin{tabular}{lc}
\hline
\hline
Corrected beam excess events                         & 77.8$\pm$8.9   \\
$\bar{\nu}_\mu+^{12}C\rightarrow\mu^++^{12}B_{g.s.}$ &  2.7$\pm$0.5   \\
accidental $e^+$ background                          &  8.2$\pm$0.8   \\
\hline
$\nu_\mu+^{12}C\rightarrow\mu^-+^{12}N_{g.s.}$       & 66.9$\pm$9.0   \\
Efficiency                                           & $16.3\pm1.2\%$ \\
$\nu_\mu$ flux ($E_\nu>$ 123.1 MeV)    & 2.03$\times10^{12}$ cm$^{-2}$ \\
$\langle\sigma\rangle$ measured  & $(5.6\pm0.8\pm1.0)\times10^{-41}$ cm$^2$ \\
\hline
\hline
$\langle\sigma\rangle$ theory      &   \\
\hline
Engel {\it et al.} \cite{Engel96} & $6.4\times10^{-41}$ cm$^2$ \\
Kolbe {\it et al.} \cite{Kolbe99} & $7.0\times10^{-41}$ cm$^2$ \\
Volpe {\it et al.} \cite{Volpe00} & $6.5\times10^{-41}$ cm$^2$ \\
Hayes and Towner \cite{Hayes00}    & $5.6\times10^{-41}$ cm$^2$ \\
\hline
\end{tabular}
\label{ta:gstate}
\end{table}
The efficiency shown includes the efficiency for muons to have 
more than 17 PMT hits as well as the electron, muon and 
beta efficiencies given in Table \ref{ta:electron}, Table \ref{ta:muon} 
and Table \ref{ta:betaeff}, respectively.
The flux-averaged cross section is 
$\langle\sigma\rangle=(5.6\pm0.8\pm1.0)\times10^{-41}$ cm$^2$, 
where the first error is statistical and the second systematic.  
The two dominant sources of systematic error are the neutrino flux (15\%) 
discussed in Section \ref{sec:source} and the effective fiducial 
volume (6\%) discussed in Section \ref{sec:analysis}.  

\begin{figure}
\centerline{\psfig{figure=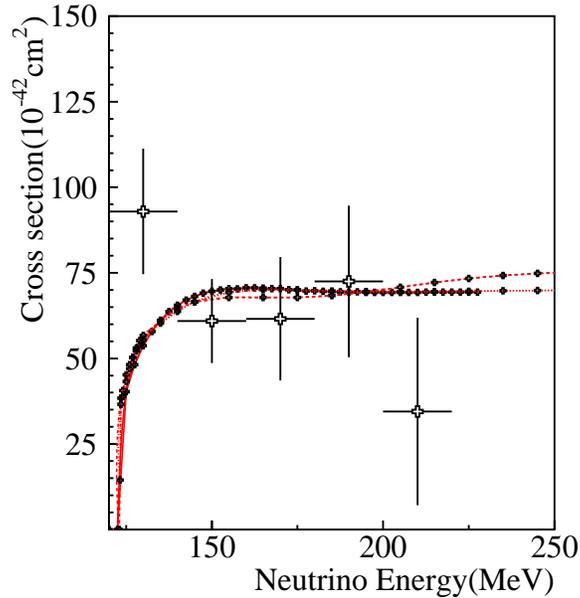,height=3.5in}}
\caption{The measured cross section for 
the process $^{12}C(\nu_\mu,\mu^-)^{12}N_{g.s.}$ compared with three 
theoretical calculations obtained from Ref. \cite{Engel96}.}
\label{fig:gs_xsec}
\end{figure}

For this reaction to the $^{12}N$ ground state it is also 
straightforward to measure the energy dependence of the cross section.  
The recoil energy of the nucleus is small and 
to a very good approximation, $E_\nu=m_\mu c^2+T_\mu+17.7$ MeV, 
where $m_\mu$ is the muon mass, 
$T_\mu$ is the kinetic energy of the muon, and 17.7 MeV arises from 
the $Q$ value of the reaction and the nuclear recoil.  
Figure  \ref{fig:gs_xsec} compares the measured cross section 
as a function of $E_\nu$ with three theoretical 
calculations obtained from Ref. \cite{Engel96}.  
The agreement is reasonable within the limited statistics.

There is little disagreement\cite{Engel96} in the predicted 
cross section for this exclusive process, 
as it is fixed by measured values of closely related 
electroweak transition probabilities.  
The differences that exist among various calculations 
result from different models for the dependence of 
various elements of the transition probability on the momentum transfer. 
Thus, as shown in Fig. \ref{fig:gs_xsec}, the differences 
among shell model approaches, an RPA calculation, and 
an ``elementary particle'' model all agree for 
$E_\nu$ up to 160 MeV and differ only by about 10\% at 250 MeV.

Table \ref{ta:gstate} also shows three more recent 
calculations \cite{Kolbe99,Volpe00,Hayes00} of the cross section 
for this exclusive process.
The focus of these papers, however, was on obtaining 
a satisfactory overall description of neutrino reactions on carbon, 
including the inclusive cross section and of $\mu^-$ capture.

\section{The Inclusive Reaction}
\label{sec:inclusive}

Most of the inclusive beam-excess events arise 
from the reaction $^{12}C(\nu_\mu,\mu^-)X$, 
but approximately 10\% are due to other sources.  
Table \ref{ta:inclusive} shows the number of beam-excess events, 
the calculated backgrounds, the efficiency, $\nu_\mu$ flux, and 
the flux-averaged cross section for this process.  
\begin{table}
\centering
\caption{Beam-excess events, background, efficiency, neutrino flux and 
flux-averaged cross section for the inclusive 
reaction $^{12}C(\nu_\mu,\mu^-)^{12}X$.}
\begin{tabular}{lc}
\hline
\hline
Corrected beam excess events             &         2464$\pm$50          \\
$\bar{\nu}_\mu+p\rightarrow\mu^++n$      &          217$\pm$35          \\
$\bar{\nu}_\mu+^{12}C\rightarrow\mu^++X$ &           71$\pm$35          \\
$\nu_\mu+^{13}C\rightarrow\mu^-+X$       &           24$\pm$12          \\
\hline
$\nu_\mu+^{12}C\rightarrow\mu^-+X$       &         2152$\pm$56          \\
Efficiency                               &      $(27.7\pm1.9) \%$       \\
$\nu_\mu$ flux ($E_\nu >$ 123.1 MeV)     & 2.03$\times10^{12}$ cm$^{-2}$\\
$\langle\sigma\rangle$ measured  & $(10.6\pm0.3\pm1.8)\times10^{-40}$ cm$^2$ \\
\hline
\hline
Theory    &      \\
\hline
Kolbe {\it et al.} \cite{Kolbe99}       &  $17.5\times10^{-40}$ cm$^2$ \\
Volpe {\it et al.} \cite{Volpe00}       &  $15.2\times10^{-40}$ cm$^2$ \\
Hayes and Towner \cite{Hayes00}          &  $13.8\times10^{-40}$ cm$^2$ \\
\hline
\end{tabular}
\label{ta:inclusive}
\end{table}
The efficiency shown includes the efficiency for muons to have 
more than 17 PMT hits as well as the electron and muon efficiencies 
given in Tables \ref{ta:electron} and \ref{ta:muon}.
The backgrounds arising from the $\bar{\nu}_\mu$ component 
of the decay-in-flight beam are small, primarily 
because the high-energy $\bar{\nu}_\mu$ flux is approximately 
a factor of 10 lower than the corresponding $\nu_\mu$ flux.  
The largest background arises from the 
process $\bar{\nu}_\mu+p\rightarrow\mu^++n$.  
The cross section is well known and the uncertainty in this process 
is mainly due to the 15\% uncertainty in the $\bar{\nu}_\mu$ flux.  
A much smaller but less well understood background arises 
from the process $^{12}C(\bar{\nu}_\mu,\mu^+)X$.  
Plausibly, as observed for the process $^{12}C(\nu_\mu,\mu^-)X$, 
the cross section might be expected to be approximately 60\% 
of that given by a recent CRPA calculation\cite{Kolbe95}. 
We use this reduced cross section in calculating this background 
but assign a large error to reflect the uncertainty in the cross section.  
An even smaller background arises from the 1.1\% $^{13}C$ component 
of the scintillator.  
For the process $^{13}C(\nu_\mu,\mu^-)X$ we use a Fermi gas model 
calculation and assign a 50\% uncertainty.

The measured flux-averaged cross section for the inclusive reaction 
$^{12}C(\nu_\mu,\mu^-)X$ 
is $\langle\sigma\rangle=(10.6\pm0.3\pm1.8)\times10^{-40}$ cm$^2$,
where the first error is statistical and the second systematic.  
The mean energy of the neutrino flux above threshold is 156 MeV.  

Table \ref{ta:inclusive} shows for comparison results of 
three recent theoretical calculations of the cross 
section \cite{Kolbe99,Volpe00,Hayes00}.
The shell model calculation of Hayes and Towner \cite{Hayes00} provides 
the best agreement with our measurement but all three predict 
a higher cross section than observed.
These calculations used the $\nu_\mu$ energy distribution calculated by 
LSND for the 1994 beam stop.
The shape of the calculated energy distribution, however, shows only 
small yearly variations that would produce shifts of a few percent in 
the expected cross section for the different years.
Averaged over all the data the calculated shape is the same 
as that calculated for 1994.

The systematic error is due almost entirely to the uncertainty 
in the $\nu_\mu$ flux.  
To determine this value, the inputs to the neutrino beam 
Monte Carlo program were varied within their estimated uncertainties.  
The resulting variation in both the magnitude and 
the shape of the $\nu_\mu$ flux above muon production threshold 
results in a 15\% uncertainty in the inclusive cross section.

The beam dump configuration was substantially modified 
after the 1995 run as discussed in Section \ref{sec:source}.
The two beam dump geometries are very different.
Also the flux for 1994-1995 depends primarily on $\pi^+$ production 
from water while the flux for 1996-1998 arises primarily 
from $\pi^+$ production on high-Z materials.
The resulting $\nu_\mu$ DIF flux per incident proton 
with the modified configuration was only one-half 
of that obtained with the water target used in 1994-1995.
Thus a comparison of the cross sections measured 
with the two beam dump configurations provides a check on 
the systematics of the $\nu_\mu$ flux simulation.
The cross section measured with the APT beam dump 
in 1996-1998, ($9.5\pm0.4(stat.))\times10^{-40}$ cm$^2$, was lower 
than that measured with the water target setup 
in 1994-1995, ($11.4\pm0.4(stat.))\times10^{-40}$ cm$^2$, 
by a little more than one standard deviation.
The statistical uncertainties of these measurements 
are small compared to the systematic uncertainties. 
The systematic errors for the two beam dump configurations are partially 
correlated but each has an uncorrelated systematic error of about 10\%.
The cross sections for the two beam dump configurations differ 
by $1.9\times10^{-40}$ cm$^2$ compared to an expected uncertainty 
of $1.5\times10^{-40}$ cm$^2$.
Thus results for the two beam dump configurations are consistent 
within errors.

\begin{figure}
\centerline{\psfig{figure=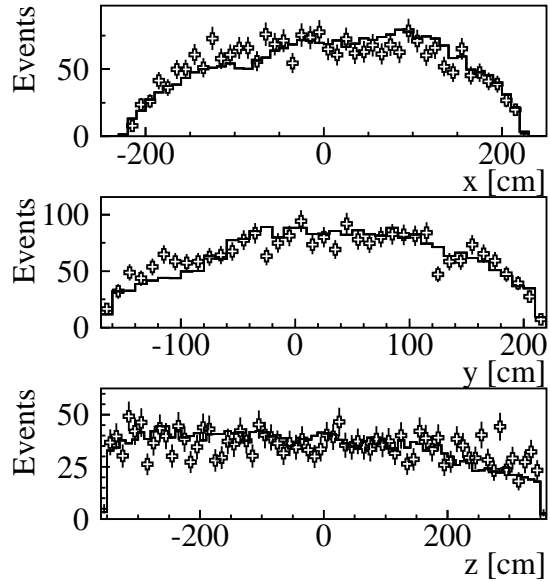,height=3.5in}}
\caption{The spatial distribution of the electron for beam-excess events 
compared with expectation (solid line) from the inclusive 
reaction $^{12}C(\nu_\mu,\mu^-)X$.}
\label{fig:xyz}
\end{figure}

The spatial distribution of the beam-excess electrons is shown 
in Fig. \ref{fig:xyz}.  
There is a clear enhancement of events at high $x$ and high $y$ 
due to the variation of the $\nu_\mu$ flux over the detector.  
The good agreement with expectation shows that this spatial distribution 
is well modeled by the beam simulation program.

\begin{figure}
\centerline{\psfig{figure=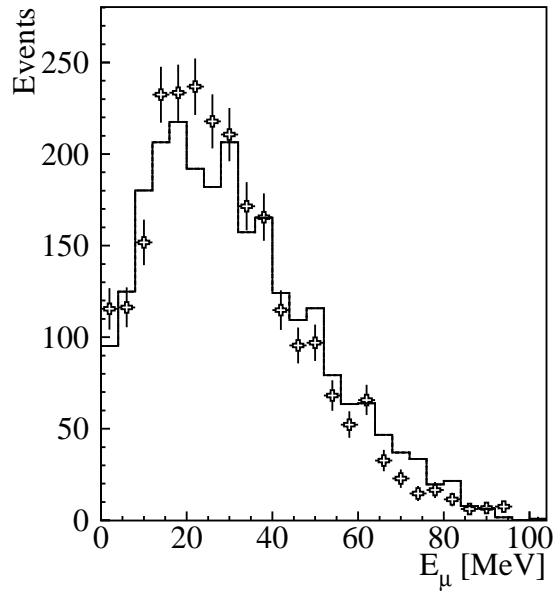,height=3.5in}}
\caption{The observed and expected (histogram) distributions of the muon 
kinetic energy, $E_\mu$, for the inclusive decay-in-flight sample.
The expected distribution has been normalized to the data.}
\label{fig:emu}
\end{figure}

For the reaction $^{12}C(\nu_\mu,\mu^-)X$, the detector charge $Q_\mu$, 
measured in photoelectrons, arises mostly from the $\mu^-$ but includes 
contributions from other particles in the reaction 
such as protons and $\gamma$'s.  
The muon kinetic energy distribution obtained 
from the $Q_\mu$ distribution is shown in Fig. \ref{fig:emu}.
According to the CRPA calculation \cite{Kolbe95}, 
84.5$\%$ (20.4$\%$) of the inclusive $^{12}C(\nu_\mu,\mu^-)X$ events 
have proton ($\gamma$) emission from the decay of the N excited states.
The average energies of these protons and $\gamma$s are 
estimated to be 9 and 4 MeV.
We used the calculation of Ref. \cite{Kolbe95} to determine proton 
and $\gamma$ energy distributions and LSNDMC to 
determine the number of photoelectrons produced.  
The averaged contribution of these particles to the measured electron 
equivalent energy of the muons is estimated to be 2.2 MeV.
Protons produce less scintillation light than electrons 
due to saturation effects.  
The uncertainty in the saturation effect is the primary source 
of uncertainty in the muon and proton energy determination.  
The average contribution to $Q_\mu$ from particles other than 
the muon is approximately 12\%.
The expected muon energy spectrum in Fig. \ref{fig:emu} is obtained
from a recent CRPA calculation and 
includes the contribution from particles other than muon.
There is reasonable agreement.  
However, given the uncertainties in the shape of the $\nu_\mu$ 
energy spectrum, in the modeling of the energy from nuclear breakup 
and in the muon and proton energy calibration, 
we do not try to extract any information on the energy dependence 
of the cross section for the reaction $^{12}C(\nu_\mu,\mu^-)X$.

\begin{figure}
\centerline{\psfig{figure=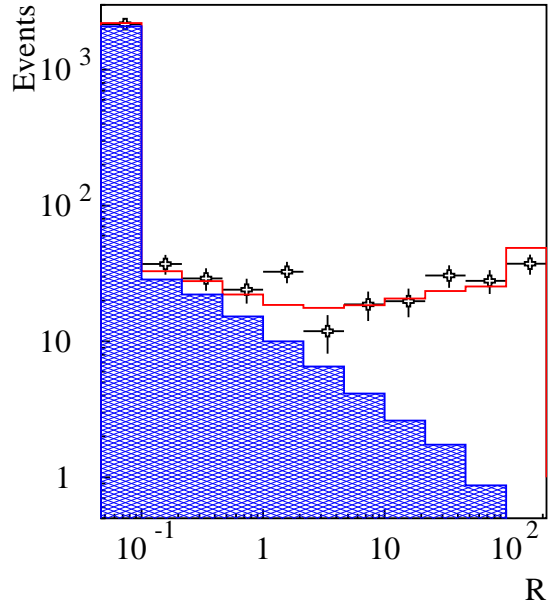,height=3.5in}}
\caption{The observed distribution of the $\gamma$ likelihood ratio R 
for the inclusive decay-in-flight sample.
Shown for comparison are the best fit (solid line) combination of 
the correlated distribution and uncorrelated distribution to the data. 
The best fit has a (11.6$\pm$1.1)$\%$ correlated component.
The shaded region shows the uncorrelated contribution.}
\label{fig:rplot}
\end{figure}

Further information on the inclusive sample can be obtained 
by measuring the fraction of the events with an associated neutron.  
The presence of a neutron is established by detection 
of the $\gamma$ ray from the neutron's capture on 
a proton in the detector via the reaction $n+p\rightarrow d+\gamma$.  
A detailed discussion on the procedure used to detect these $\gamma$'s 
can be found in Ref. \cite{LSND_osc01}. 
The distribution of the likelihood ratio $R$ for 
correlated $\gamma$'s from neutron capture is very different 
from that for uncorrelated (accidental) $\gamma$'s.  
The measured $R$ distribution for the inclusive sample, shown 
in Fig. \ref{fig:rplot}, was fit to a mixture of these two 
possible $\gamma$ sources 
to determine the fraction of events with associated neutrons.  
The best fit, also shown in the figure, corresponds to a fraction 
of events with a neutron of $(11.6\pm1.1)$\%, 
where the error includes systematic uncertainties.

The two largest backgrounds shown in Table \ref{ta:inclusive} arise 
from the $\bar{\nu}_\mu$ component of the beam, 
and almost all of these events should have an associated neutron.  
In contrast, most of the events arising from $\nu_\mu$ 
interactions will not have an associated neutron.  
A CRPA calculation predicts that 79\% of the events 
from the reaction $^{12}C(\bar{\nu}_\mu,\mu^+)X$ will have 
an associated neutron compared to only 5.9\% 
for the reaction $^{12}C(\nu_\mu,\mu^-)X$\cite{Kolbe95}.

Table \ref{ta:component} shows the measured component with 
an associated neutron for the beam excess sample and 
the calculated backgrounds from $\bar{\nu}_\mu$ interactions. 
The resulting number for the $\nu_\mu$ carbon sample ($1.0\pm2.4$\%) 
is lower than the CRPA prediction of 5.9\%.  
\begin{table}
\centering
\caption{The expected and observed numbers of events with associated 
neutrons and the calculated background from $\bar{\nu}_\mu$ reactions.}
\begin{tabular}{cccc}
\hline
            & Events from              &   Fraction   &  Events with  \\
Source      & Table \ref{ta:inclusive} & with neutron &   neutron     \\
\hline
Beam Excess                      & 2464 & $(11.6\pm1.1)\%$ & 286$\pm$27 \\
$\bar{\nu}_\mu p\rightarrow\mu^+n$  &  214    &   $100\%$  & 214$\pm$35 \\
$\bar{\nu}_\mu C\rightarrow\mu^+nX$ &   64    &    $79\%$  &  51$\pm$26 \\
\hline
$\nu_\mu C\rightarrow\mu^-X$        &  2181 & $1.0\pm2.4\%$ & 21$\pm$52 \\
\hline
\end{tabular}
\label{ta:component}
\end{table}
The CRPA prediction is for the $\nu_\mu$ spectrum used in our earlier 
analysis \cite{LSND97b}, which is slightly harder than that used in the present analysis as discussed in Section \ref{sec:source}.
The observed number of events with neutrons also rules out 
a $\bar{\nu}_\mu$ flux much bigger than that calculated 
by the beam Monte Carlo simulation.

\section{Conclusions}
\label{sec:conclusion}

The exclusive process $^{12}C(\nu_\mu,\mu^-)^{12}N_{g.s.}$ has been 
measured with a clean sample of $66.9\pm9.1$ events 
for which the $\mu^-$, the decay $e^-$, and the $e^+$ 
from the $\beta$ decay of the $^{12}N_{g.s.}$ are detected.  
For this process the theoretical cross section calculations are 
very reliable.  
The flux averaged cross section is measured to be 
$(5.6\pm0.8\pm1.0)\times10^{-41}$ cm$^2$ 
in good agreement with theoretical expectations.  
From comparison of this cross section with the cross section 
for the inclusive process $^{12}C(\nu_{\mu},\mu^-)X$ 
we obtain a flux-averaged branching ratio of $(5.3\pm0.8\pm0.5)\%$.

The inclusive process $^{12}C(\nu_\mu,\mu^-)X$ has also been measured.  
There are model-dependent uncertainties in the theoretical cross section 
calculation for this process 
that are not present for the $^{12}N_{g.s.}$ cross section.  
The measured flux-averaged cross section, 
$(10.6\pm0.3\pm1.8)\times10^{-40}$ cm$^2$, is smaller than theoretical 
expectations \cite{Kolbe99,Volpe00,Hayes00}.
It is in better agreement with the shell model calculation of Hayes and 
Towner \cite{Hayes00} than with the CRPA calculation 
of Kolbe {\it et al.} \cite{Kolbe99}.
The mean energy of the neutrino flux above threshold is 156 MeV.  
The measured distribution of the muon energy 
(including contributions from other particles such as protons 
and $\gamma$'s) agrees within errors with the CRPA calculation.  
The fraction of events with associated neutrons was measured 
to be $(1.0\pm2.4)$\%. 
This is lower than the CRPA calculation of 5.9\%.

\label{sec:ack}

{\it Acknowledgments} 
This work was conducted under the auspices of the US Department 
of Energy, supported in part by funds provided by the University 
of California for the conduct of discretionary research 
by Los Alamos National Laboratory.
This work was also supported by the National Science Foundation.
We are particularly grateful for the extra effort that was made 
by these organizations to provide funds for running the accelerator 
at the end of the data taking period in 1995.
It is pleasing that a number of undergraduate students from participating 
institutions were able to contribute significantly to the experiment.

\end{document}